\documentclass[twocolumn,floats,showpacs,superscriptaddress,pre]{revtex4}
\usepackage{amsmath,amssymb,bm}
\usepackage{graphics}
\usepackage{epsfig}
\usepackage{graphicx}

\begin{document}

\title{ On Statistical Distribution for Adiabatically Isolated Body\\
}
\author{Natalia Gorobey}
\affiliation{Peter the Great Saint Petersburg Polytechnic University, Polytekhnicheskaya
29, 195251, St. Petersburg, Russia}
\author{Alexander Lukyanenko}
\email{alex.lukyan@mail.ru}
\affiliation{Peter the Great Saint Petersburg Polytechnic University, Polytekhnicheskaya
29, 195251, St. Petersburg, Russia}
\author{A. V. Goltsev}
\affiliation{Ioffe Physical- Technical Institute, Polytekhnicheskaya 26, 195251, St.
Petersburg, Russia}

\begin{abstract}
The statistical distribution for the case of an adiabatically isolated body
was obtained in the framework of covariant quantum theory and Wick's
rotation in the complex time plane. The covariant formulation of the
mechanics of an isolated system lies in the rejection of absolute time and
the introduction of proper time as an independent dynamic variable. The
equation of motion of proper time is the law of conservation of energy. In
this case, the energy of an isolated system is an external parameter for the
modified distribution instead of temperature.
\end{abstract}

\maketitle







\section{\textbf{INTRODUCTION}}

Statistical Boltzmann distribution with density

\begin{equation}
f\left( q,p\right) =A\exp \left[ -\frac{H\left( q,p\right) }{k_{B}T}\right]
\label{1}
\end{equation}%
for an adiabatically isolated body has an approximate character \cite{Landau}%
. It can be used to calculate the statistical characteristics of small
subsystems of a macroscopic body, but when calculating such quantities as
the total energy and its fluctuations, it leads to obviously incorrect
results. Although the calculation errors in this case are not large compared
to the quantities themselves and may not be taken into account in the
macroscopic approximation, the search for an adequate replacement of
distribution Eq.(\ref{1}) for an adiabatically isolated body is of
fundamental importance.

In this search, we use the formulation of the laws of statistics in quantum
mechanics \cite{Fey},\cite{FeyHibbs} and the transparent analogy with
quantum dynamics itself available here. It lies in the analogy between the
parabolic equation for the density matrix,

\begin{equation}
\frac{\partial \widehat{\rho }}{\partial \beta }=-\widehat{H}\widehat{\rho },
\label{2}
\end{equation}%
where $\beta =1/k_{B}T$, and the Schr\"{o}dinger equation for the evolution
operator,

\begin{equation}
i\hbar \frac{\partial \widehat{U}}{\partial t}=\widehat{H}\widehat{U}.
\label{3}
\end{equation}%
It can be seen that the evolution operator

\begin{equation}
\widehat{U}=\exp \left( \frac{1}{i\hbar }\widehat{H}t\right)  \label{4}
\end{equation}%
turns into a density operator (matrix) by a simple replacement

\begin{equation}
t\rightarrow i\hbar \beta ,  \label{5}
\end{equation}%
i.e., as a result of the transition to imaginary time. In turn, the
classical Boltzmann distribution Eq. (\ref{1}) can be obtained from the
mixed coordinate-momentum representation of the logarithm of the density
operator due to the relation

\begin{equation}
\beta H\left( q,p\right) =\beta \left\langle p\right\vert \widehat{H}%
\left\vert q\right\rangle =-\left\langle p\right\vert \ln \widehat{\rho }%
\left\vert q\right\rangle ,  \label{6}
\end{equation}%
which is valid if we agree to place the momentum operators in $\widehat{H}$
to the right of the coordinates. In \cite{FSS}, the definition of the
partition function of an adiabatically isolated body in the form of a trace
of the density operator in the modified quantum theory was proposed. The
modification consists in applying the rules of covariant quantization \cite%
{Frad} to a dynamical system with reparametrization time invariance. In such
a system, proper time is naturally introduced as an additional dynamic
variable. Its equation of motion is the law of conservation of energy of the
system. In the transition to statistical mechanics, as a result of
replacement Eq. (\ref{5}), we obtain a modified partition function as a
function of the energy of an isolated system. In this paper, we will go
further in the modification of statistical mechanics, and using relation Eq.
(\ref{6}), we will determine the analogue of distribution Eq. (\ref{1}) for
an isolated body.

The first section formulates the dynamics of an isolated mechanical system
in terms of proper time. The second section proposes a quantum version of
the modified dynamics and a modified statistical distribution.

\section{MECHANICS OF AN ISOLATED BODY}

Mechanical laws can be formulated from the outset in a form that explicitly
takes into account the complete isolation of the system under consideration
from its environment. First of all, it concerns time. Instead of the
absolute Newton time $t$, we introduce an arbitrary parameter $\tau \in %
\left[ 0,1\right] $ and a new dynamic variable $s\left( \tau \right) $:

\begin{equation}
s\left( \tau \right) =\frac{dt}{d\tau }.  \label{7}
\end{equation}%
Accordingly, we write the action integral in the form

\begin{equation}
I\left[ q,\overset{\cdot }{q},\overset{\cdot }{s}\right] =\int_{0}^{1}d\tau
\overset{\cdot }{s}L\left( q,\frac{\overset{\cdot }{q}}{\overset{\cdot }{s}}%
\right) .  \label{8}
\end{equation}%
Obviously, action Eq. (\ref{8}) is invariant under an arbitrary change of
the parameter $\tau $. The new dynamic variable $s\left( \tau \right) $ will
play the role of the proper time of the system, if we take into account its
equation of motion,

\begin{equation}
\frac{\delta I}{\delta s}=\frac{dW}{d\tau }=0,  \label{9}
\end{equation}%
where

\begin{equation}
W\equiv \frac{\partial L\left( q,q^{\prime }\right) }{\partial q_{k}^{\prime
}}q_{k}^{\prime }-L,  \label{10}
\end{equation}%
is the energy, $k=1,2,\ldots ,K,K$ - number of degrees of freedom of the
system. Here the prime denotes the derivative with respect to proper time.
Indeed, as we know, the first integral of motion

\begin{equation}
W=E  \label{11}
\end{equation}%
allows you to determine your own time by the movement of dynamic variables $%
q_{k}\left( s\right) $. It remains only to fix the surface of constant
energy Eq. (\ref{11}) as an additional condition in the action integral:

\begin{equation}
\widetilde{L}\left( q,q^{\prime }\right) =L\left( q,q^{\prime }\right)
+\lambda ^{\prime }W.  \label{12}
\end{equation}%
We have discarded the total derivative $\lambda ^{\prime }E$ in Eq. (\ref{12}%
). Note that the arbitrariness in the choice of the parameter $\tau $ is now
expressed in the arbitrariness of the interval of motion in proper time $%
s\in \left[ 0,C\right] $, which is removed only by the additional condition
Eq. (\ref{11}).

As a next step, we find the Hamilton function of the modified system. First,
we define the canonical momenta:

\begin{equation}
\widetilde{p}_{k}=m_{kl}q_{l}^{\prime }\left( 1-\lambda ^{\prime }\right) ,
\label{13}
\end{equation}

\begin{equation}
\widetilde{p}_{\lambda }=W=\frac{1}{2}m_{kl}q_{k}^{\prime }q_{l}^{\prime
}+V\left( q\right) ,  \label{14}
\end{equation}%
where the stroke denotes the derivative with respect to proper time and we
specified the form of the original Lagrange function:

\begin{equation}
L\left( q,\overset{\cdot }{q}\right) =\frac{1}{2}m_{kl}\overset{\cdot }{q}%
_{k}\overset{\cdot }{q}_{l}-V\left( q\right) .  \label{15}
\end{equation}%
With the help of these equations we find speeds. At first,

\begin{equation}
q_{k}^{\prime }=\frac{m_{kl}^{-1}\widetilde{p}_{l}}{\left( 1-\lambda
^{\prime }\right) },  \label{16}
\end{equation}%
and then,

\begin{equation}
\lambda ^{\prime }=1-\frac{\sqrt{m_{kl}^{-1}\widetilde{p}_{k}\widetilde{p}%
_{l}}}{\sqrt{2\left( E-V\right) }},  \label{17}
\end{equation}%
where Eq. (\ref{11}) is taken into account. After that, we find the modified
Hamilton function:

\begin{equation}
\widetilde{H}=\sqrt{2\left( E-V\right) }\sqrt{m_{kl}^{-1}\widetilde{p}_{k}%
\widetilde{p}_{l}}+E-2V.  \label{18}
\end{equation}%
Thus, in modified mechanics, the proper time of motion of an isolated body
is an indeterminate parameter (before solving the classical equations of
motion), and the energy has a fixed value. The two-valued root in Eq. (\ref%
{18}) corresponds to two half-cycles of the system's motion between two
cusps in the region of allowed classical motion. After quantization, a root
from the elliptic operator will arise, which will need to be given meaning.

In the final step, after quantization, we return to absolute time by simply
setting $s=t$. This is acceptable, since the dynamics of an isolated body
can itself serve to measure the time $t$. There is no obstacle to this in
the new dynamics. Indeed, setting $\widetilde{p}_{k}=p_{k}$, from Eq. (\ref%
{13}) we obtain $\lambda ^{\prime }=0$. Then relation Eq. (\ref{17}) is
equivalent to Eq. (\ref{11}), and $\tilde{H}=E$, as expected for an isolated
system with energy $E$. In what follows, we will omit the "tilde" in the
designation of canonical momenta.

\section{STATISTICAL DISTRIBUTION FOR AN ISOLATED BODY}

For convenience, we set here $\hbar =k_{B}=1$. Let us represent the kernel
of the evolution operator of the modified quantum theory as a functional
integral on the phase space \cite{FadSlav}:

\begin{eqnarray}
D &\equiv &\left\langle q^{\prime \prime },C\right\vert \left. q^{\prime
},0\right\rangle =\int \prod\limits_{s}\frac{d^{K}pd^{K}q}{2\pi }\exp
\left\{ i\int_{0}^{C}ds\left[ p_{k}\overset{\cdot }{q}_{k}\right. \right.
\notag \\
&&-\left. \left. \widetilde{H}\left( q,p,E\right) \right] \right\} .
\label{19}
\end{eqnarray}%
The formal nature of the functional integral in Eq. (\ref{19}) is aggravated
by the presence of a root from the quadratic form of the momenta in the
Hamiltonian. We make the kernel more definite by turning the functional
integral Eq. (\ref{19}) into Gaussian over canonical momenta by introducing
integrals over additional variables:

\begin{eqnarray}
D &=&\int \prod\limits_{s}\frac{d^{K}pd^{K}q}{\left( 2\pi \right) ^{K}}%
\int_{-\infty }^{+\infty }\prod\limits_{s}\frac{d\chi }{2\pi }%
\int_{0}^{+\infty }\prod\limits_{s}dQ^{2}  \notag \\
&&\times \exp \left\{ i\int_{0}^{C}ds\left[ p_{k}\overset{\cdot }{q}%
_{k}+E-2V\right. \right.  \notag \\
&&\left. \left. -Q+\chi \left( 2\left( E-V\right) m_{kl}^{-1}\widetilde{p}%
_{k}\widetilde{p}_{l}-Q^{2}\right) \right] \right\} .  \label{20}
\end{eqnarray}%
It is easy to see that integration first over the variable $\chi $ and then
over the variable $Q^{2}$ leads to the original integral Eq. (\ref{19}).
However, now integration over canonical momenta can be done in the first
place, and this will be the definition of the square root of the kinetic
energy in the modified quantum theory. In order to avoid cumbersome
expressions, we agree in what follows to omit numerical factors of type $%
2\pi $. As a result, we get

\begin{eqnarray}
D &=&\int \prod\limits_{s}d^{K}q\int_{-\infty }^{+\infty }\prod\limits_{s}%
\frac{d\chi }{\left[ 2i\chi \left( E-V\right) \right] ^{K/2}}  \notag \\
&&\times \int_{-\infty }^{+\infty }\prod\limits_{s}QdQ\exp \left\{
i\int_{0}^{C}ds\left[ -Q\right. \right.   \notag \\
&&-\left. \left. \chi Q^{2}-\frac{1}{4\chi }\frac{m_{kl}q_{k}^{\prime
}q_{l}^{\prime }}{2\left( E-V\right) }+E-2V\right] \right\} ..  \label{21}
\end{eqnarray}%
To see what the additional integrals in Eq. (\ref{21}) now lead to, we
calculate the next integral over the variable $Q$ in the semiclassical
approximation \cite{Fed}:

\begin{equation}
D\cong \int \prod\limits_{s}\frac{d^{K}q}{\sqrt{E-V}}\Delta \left[ W-E\right]
\exp \left[ i\int_{0}^{C}ds\left( E-2V\right) \right] .  \label{22}
\end{equation}%
Here
%
%

\begin{eqnarray}
\Delta [ W-E ] \equiv 
\mathrm{Re} \int_{0}^{\infty} \prod \limits_{s}\left( i\eta \right)^{\left( K-1 \right) /2}d\eta
\nonumber \\
\times \exp \left[ i\int_{0}^{C}ds\frac{\eta }{4}\left( \frac{1}{2} m_{kl} q_{k}^{\prime } q_{l}^{\prime}+V-E\right) \right] ,
\label{23}
\end{eqnarray}
where $\eta =1/\chi(E-V) $ is the new integration variable. As
expected, functional Eq. (\ref{23}) limits integral Eq. (\ref{22}) to
trajectories that lie entirely on the surface of constant energy $E$. For
such trajectories, the exponent under the integral sign in Eq. (\ref{22})
exactly coincides with the initial action of the system.

Thus, the presence of the square root of the quadratic pulse shape in the
modified Hamilton function Eq. (\ref{18}) is the source of the generalized $%
\delta $ -function Eq. (\ref{23}) in the modified propagator Eq. (\ref{22}).
This is just what is needed to solve another problem of covariant quantum
theory, the problem of time $C$. A similar problem of determining the
propagator of a relativistic particle within the framework of the rules of
covariant quantum theory \cite{Frad} is completed by additional integration
over proper time within $0\leq C<\infty $ with the trivial measure $\mu =1$
\cite{Gov}. In our problem, the proper time on an arbitrary trajectory $q(s)$
is determined two-valuedly by two directions of motion, so we should
integrate along the full axis $-\infty <C<\infty $. It should be borne in
mind that the time of movement on the trajectory $q(s)$ is a functional of
the trajectory: $C=C[q(s)]$. Therefore, an additional integral over proper
time in the propagator should be placed before the final summation over all
admissible trajectories. Now the purpose of the proposed modification of
mechanics is clear: the generalized $\delta $-function Eq. (\ref{23}) will
allow removing additional integration over proper time in covariant quantum
theory and giving the propagator a dynamic meaning, and after the transition
to imaginary proper time, a statistical meaning to the density matrix. The
resulting proper time, and in statistical mechanics, the reciprocal
temperature of an adiabatically isolated body is random variable in this
theory. Thus, we obtain the following representation of the density matrix
of an adiabatically isolated body:

\begin{eqnarray}
\widetilde{\rho }\left( q^{\prime \prime },q^{\prime }\right) &=&\int
\prod\limits_{s}\frac{d^{K}q}{\sqrt{V-E}}  \notag \\
&&\times \int_{-\infty }^{+\infty }dC\Delta \left[ V-\frac{1}{2}%
m_{kl}q_{k}^{\prime }q_{l}^{\prime }-E\right]  \notag \\
&&\times \exp \left[ -\int_{0}^{C}ds\left( E-2V\right) \right] .  \label{24}
\end{eqnarray}%
From the quantum mechanical density matrix Eq. (\ref{24}), we proceed to the
modification of the classical Boltzmann distribution Eq. (\ref{1}) for an
adiabatically isolated body using relation Eq. (\ref{6}):

\begin{equation}
\widetilde{f}\left( q,p,E\right) =A\exp \left[ B\left( q,p,E\right) \right] ,
\label{25}
\end{equation}%
where

\begin{equation}
B\left( q,p,E\right) =\left\langle q\right\vert \ln \widetilde{\rho }%
\left\vert p\right\rangle .  \label{26}
\end{equation}

\section{CONCLUSIONS}

Thus, the statistical distribution for an adiabatically isolated body is
defined within the framework of the formalism of reparameterization
invariant mechanics and the corresponding covariant quantization procedure.
The indeterminate proper time parameter arising in this formalism also plays
the role of an indefinite body temperature in modified statistical
mechanics. This is how it should be in a system that is devoid of contact
with the environment and whose energy is fixed. It must also be assumed that
the nominal distribution Eq. (\ref{25}) in accordance with the ergodic
hypothesis \cite{Arn} describes the relative time that the system spends in
the vicinity of the point $\left( q_{k},p_{k}\right) $ of the phase space
when moving on the surface of constant energy $E$.

Distribution Eq. (\ref{25}) can only be used for a macroscopic body with a
large amount of mechanical energy. In this case, the continuous law of
conservation of mechanical energy Eq. (\ref{9}) is obviously substantiated
in quantum theory in the semiclassical approximation. At the same time, the
modification of the classical statistical distribution proposed here is
justified. 

The next step for studying the statistics of an isolated system is the calculation 
of the distribution density Eq. (25) for the simplest system - an ensemble of harmonic oscillators. 
This will be the subject of subsequent works.

\section{ACKNOWLEDGEMENTS}

We are thanks V.A. Franke for useful discussions.




\end{document}